\documentclass[aps,onecolumn,superscriptaddress,showkeys,showpacs,longbibliography,nofootinbib]{revtex4-2}
\usepackage[utf8]{inputenc}
\usepackage{epsfig,amsmath,amssymb}
\usepackage{bm}
\usepackage{graphicx}
\usepackage{graphicx}
\usepackage{dcolumn,color}

\usepackage[colorlinks, linkcolor=red,anchorcolor=green,citecolor=blue]{hyperref}
\usepackage[normalem]{ulem}

\newcommand{\Tr}{\mathop{\mathrm{Tr}}}
\newcommand\diag{\operatorname{diag}}

\newcommand{\bx}{\bm{x}}
\newcommand{\bk}{\bm{k}}
\newcommand{\nn}{\nonumber}

\begin{document}

\title{Cross effects in spin hydrodynamics: \protect\\
	Entropy analysis and statistical operator}
\author{Jin Hu}
\email{hu-j17@mails.tsinghua.edu.cn}
\affiliation{Department of Physics, Tsinghua University, Beijing 100084, China}

\begin{abstract}
	We revisit the construction of first-order spin hydrodynamics and find that  the constitution relations  receive the corrections from 
	the cross effects resulting from spin-orbit coupling. 
 Starting from a  routine entropy analysis, we show how to identify cross effects  and  new cross transport coefficients from the second law of thermodynamics.  Interestingly, the conventional transport coefficient heat conductivity $\kappa$  is bounded from below by the product of cross transport coefficients, which means the threshold of heat conduction is changed. With recourse to Zubarev's  non-equilibrium statistical operator, we reproduce the construction of first-order spin hydrodynamics and identification of cross effects in a more rigorous way. 
  By seeking the dispersion relations of normal modes, we find that
    these  cross  effects suppress the attenuation of sound modes and heat mode appearing in conventional hydrodynamics and also have impacts on the damping of non-hydrodynamic spin modes. 
  \end{abstract}

\maketitle

\section{Introduction} 
It has been proposed for long that the final hadrons produced in heavy-ion collisions are polarized by the strong orbital angular momentum \cite{Liang:2004ph,Voloshin:2004ha,Betz:2007kg,Becattini:2007sr} similar to the well-known physical phenomenon, Barnett effect, in the context of condensed matters \cite{Barnett:1935}. 
Motivated by the recent measurements of spin polarization of $ \Lambda $ hyperons and observed phenomena of spin alignments in the experiment of heavy-ion collisions \cite{STAR:2017ckg,Alpatov:2020iev,Adam:2019srw,Adam:2018ivw}, the evolution of spin in hot QCD plasma draws extensive attention and constant theoretical efforts are devoted to it.  Due to the huge successes made by relativistic hydrodynamics in describing the evolution of QGP (quark gluon plasma) created in the collisions, spin hydrodynamics, namely, relativistic  hydrodynamics incorporating spin degrees of freedom, is considered as  a promising framework. 
In order to  construct a  consistent theory of  relativistic spin hydrodynamics, there have been  many theoretical studies  along this line,  which  are   based on a general entropy analysis or the second law of thermodynamics \cite{Hattori:2019lfp,Fukushima:2020ucl,Li:2020eon,Gallegos:2021bzp},  quantum kinetic theory of fermions \cite{Florkowski:2017ruc,Bhadury:2020cop,Shi:2020htn,Peng:2021ago,Hu:2021pwh,Hu:2022xjn,Weickgenannt:2022zxs}, holographic  approach \cite{Hashimoto:2013bna,Garbiso:2020puw,Gallegos:2020otk}, effective action  \cite{Montenegro:2017rbu,Montenegro:2020paq} and statistical density operator \cite{Becattini:2009wh,Becattini:2012pp,Becattini:2018duy,Hu:2021lnx}. 

Starting from a routine entropy analysis, one can construct  a first order hydrodynamic theory \cite{Hattori:2019lfp,Fukushima:2020ucl}. In general, the energy momentum tensor $T^{\mu\nu}$ has an antisymmetric part  compared to conventional dissipative first order hydrodynamics. However, it is widely believed that the definitions for $T^{\mu\nu}$ and the spin tensor $S^{\lambda\mu\nu}$ is not unique and distinct definition choices are related via pseudo gauge transformation \cite{Hehl:1976vr,Becattini:2018duy,Speranza:2020ilk}. Therefore one can show that  a spin hydrodynamic theory is equivalent to a conventional hydrodynamics with spin corrections \cite{Fukushima:2020ucl, Li:2020eon}. Hereafter, we confine our discussion to the case of a non-vanishing  antisymmetric part of $T^{\mu\nu}$. In these mentioned studies, the constitutive relations and hydrodynamic motion equations in conventional sector are immune to spin correction. They act as if they do not 'feel'  the influence of spin correction. The only differences from conventional theory are that more motion equations  appear  and new dissipative quantities and transport coefficients are present in spin sector. Likewise, they have no interplay with conventional sector. As a supplement, we comment that this does not mean that the mechanism of spin-orbit coupling does not set in, which lies in the decomposition of total angular momentum tensor $\Sigma^{\lambda\mu\nu}$ into spin and orbital contributions. Accounting for this point, the relation-type motion equations of spin densities and gapped non-hydrodynamic spin modes \cite{Hattori:2019lfp} are exactly the reflection of spin-orbit coupling .

We revisit the construction of first order spin hydrodynamics and find a new source responsible for spin-orbit coupling based on the entropy analysis and statistical operator method. Our findings indicate that the constitutive relations of conventional sector also suffer from spin corrections and vice versa. There exist the cross effects between the symmetric and antisymmetric parts of the energy momentum tensor. Furthermore, these cross effects are included in spin hydrodynamic motion equations through modified constitutive relations, therefore the evolution of spin densities will be affected by them.
We note these cross effects do not appear in previous related studies \cite{Hattori:2019lfp,Fukushima:2020ucl,Hu:2021lnx}. 

We employ natural units $\hbar=k_B=c=1$. The metric tensor here is given by $g^{\mu\nu}=\diag(1,-1,-1,-1)$, while $\Delta^{\mu\nu} \equiv g^{\mu\nu}-u^\mu u^\nu$ is the projection tensor orthogonal to the four-vector fluid velocity $u^\mu$ with $u^\mu u_\mu = 1$.  In the following, the shorthand notations $\nabla^\mu \equiv \Delta^{\mu\nu}\partial_\nu$, $D\equiv u^\mu\partial_\mu$ are used as the spatial and temporal  component of derivative.
In addition, we utilize the symmetric/antisymmetric shorthand notations:
$X^{( \mu\nu ) } \equiv (X^{ \mu\nu } + X^{ \nu \mu})/2, 
X^{[ \mu\nu ] } \equiv (X^{ \mu\nu } - X^{ \nu \mu})/2, 
X^{\langle \mu\nu \rangle}\equiv
	\bigg(\frac{\Delta^{\mu}_{\alpha} \Delta^{\nu}_{\beta} 
	+ \Delta^{\nu}_{\alpha} \Delta^{\mu}_{\beta}}{2}
	 - \frac{\Delta^{\mu\nu} \Delta_{\alpha\beta}}{3}\bigg)X^{\alpha\beta}$.\\

\section{Entropy analysis} 
 All macroscopic conservation laws relevant for the evolution of spinful fluids read
\begin{align}
\label{con}
\partial_\mu T^{\mu\nu} =0,\quad 
\partial_\mu N^{\mu} = 0,\quad
\partial_\lambda \Sigma^{\lambda\mu\nu} =0 
\, .
\end{align}
Here $N^{\mu}, \Sigma^{\lambda\mu\nu}$ are the conserved current and total angular momentum tensor.
These three equations express the conservation of the energy-momentum tensor, conserved current and total angular momentum tensor respectively.
The rank three tensor $\Sigma^{\lambda\mu\nu}$ is often canonically decomposed into two distinct parts  
$ \Sigma^{\lambda\mu\nu}
= (x^\mu T^{\lambda\nu}  - x^\nu T^{\lambda\mu} ) +  S^{\lambda\mu\nu} $, where $ S^{\lambda\mu\nu} = - S^{\lambda\nu\mu} $.
Then the last equation of Eq.(\ref{con}) can be also written as
\begin{align}
\label{Scon}
\partial_\lambda S^{\lambda\mu\nu} = T^{\nu\mu}-T^{\mu\nu}\, ,
\end{align}
where $ T^{\mu\nu}$ has both symmetric and antisymmetric nonzero
components: $ T^{\mu\nu} \equiv T^{(\mu\nu)} + T^{[\mu\nu]} $.  


Note that we have modified thermodynamic relation in equilibrium
\begin{align}
\label{eq:FirstLaw}
Ts+\mu n = e+P -\frac{1}{2}\omega_{\mu\nu} S^{\mu\nu},
 \end{align}
where $ T $, $ s $, $ \mu $, $ n $, $e  $, and $P  $ denote
the local temperature, entropy density, chemical potential, conserved charge density, energy density, and static pressure, respectively. In addition,  we introduce a ``spin potential'' $\omega_{\mu\nu} $  conjugate to the spin density $ S^{\mu\nu} $ and it is counted as $O(\partial^1)$ in derivative  counting scheme following the same prescription used in \cite{Hattori:2019lfp}.
Note $\omega_{\mu\nu} $ and $S_{\mu\nu}$ are both antisymmetric.

  Applying the derivative expansion and assuming the parity symmetry of the system, the constitutive relations can be organized  as
\begin{align}
\label{t}
&
T^{\mu\nu} = e u^\mu u^\nu - P \Delta^{\mu\nu} + T^{\mu\nu}_{(1)}
\, ,
\\
\label{n}
	&
	N^{\mu} = n u^\mu + j^{\mu}
	\, ,
\\
\label{s}
&
S^{\mu\alpha\beta} =  u^\mu S^{\alpha\beta} +S^{\mu\alpha\beta} _{(1)}
\, \\
&s^{\mu}=su^\mu+j^\mu_s,
\end{align}
and $T^{\mu\nu}_{(1)}$ is split into two parts
\begin{eqnarray}
&&
\label{ts}
T^{(\mu\nu)}_{(1)} = 2h^{(\mu}u^{\nu)}+\pi^{\mu\nu}+\Pi\Delta^{\mu\nu}, \\
\label{ta}
&& T^{[\mu\nu]}_{(1)} = 2q^{[\mu}u^{\nu]}+\tau^{\mu\nu} ,
\end{eqnarray}
where $T^{\mu\nu}_{(1)}, S^{\mu\alpha\beta} _{(1)}$ denote the first order correction to the energy momentum tensor and spin tensor, $j^\mu, j^\mu_s$ are the charge diffusion and entropy fluxes, $\pi^{\mu\nu}$ and $\Pi$ denote shear stress tensor and bulk viscous pressure, and $h^\mu$ is heat flow. At the meanwhile, $\tau^{\mu\nu}$ and $q^\mu$ are the counterparts of $\pi^{\mu\nu}$ and $h^\mu$ in the antisymmetric sector.  We also require that  $ j_s^\mu u_\mu =j^\mu u_\mu =h^\mu u_\mu = q^\mu u_\mu= \tau^{\mu\nu} u_\nu = \pi^{\mu\nu}u_\nu= 0 $.

When considering nonzero spin density and spin potential,  the entropy current is assumed to have the familiar form used in \cite{Israel:1979wp},
\begin{align}
\label{eqs1}
s^\mu
& = \frac{u_\nu}{T}T^{\mu\nu} +\frac{p}{T}u^{\mu} s -\frac{\mu}{T}j^{\mu} -\frac{1}{2} \frac{1}{T}\omega_{\alpha\beta}S^{\alpha\beta}u^{\mu}+O(\partial^2) 
\nn
\\
&=su^\mu + \frac{u_\nu}{T}T^{\mu\nu}_{(1)}-\frac{\mu}{T}j^{\mu} + O(\partial^2) .
\end{align} 
Temporarily, this is treated as an ansatz and it will be verified in the next section that this prescription is consistent with the non-equilibrium entropy current employed in the statistical operator.


Combined with the  thermodynamic relation and hydrodynamic equations, one can derive the divergence of the entropy current,
\begin{align}
\label{entropy}
&\partial_\mu s^\mu
 = \frac{n}{e+p}h^{\mu\prime}\nabla_\mu \frac{\mu}{T} 
	+\frac{\pi^{\mu\nu}}{T}\partial_{\langle\mu} u_{\nu\rangle}+\frac{\Pi}{T}\theta
\nn
\\
&\
 + q^\mu \Big(-\frac{u\cdot \partial}{T}u_\mu+\partial_\mu\frac{1}{T}+\frac{2\omega_{\mu\nu}u^\nu}{T}\Big)
\nn
\\
&\
+\tau^{\mu\nu} \Big[\frac{1}{2}\Delta_{\mu\rho}\Delta_{\nu\sigma}\Big(\partial^\rho \frac{u^\sigma}{T}-\partial^\sigma \frac{u^\rho}{T}\Big)+\frac{\omega_{\mu\nu}}{T}\Big]\geq 0,
\end{align}
where the non-negative sign follows from the second law of  thermodynamics,
and we use the notations  $\theta\equiv\partial_\mu u^\mu, h^{\mu\prime}\equiv h^\mu-\frac{e+P}{n}j^{\mu}$. It is clearly shown that  in equilibrium
\begin{align}
\label{wij}
&\omega_{\mu\nu}=-\frac{T}{2}\omega^{th}_{\mu\nu}, \quad \omega^{th}_{\mu\nu}\equiv\Delta_{\mu\rho}\Delta_{\nu\sigma}(\partial^\rho \frac{u^\sigma}{T}-\partial^\sigma \frac{u^\rho}{T}),\\
\label{w0i}
&\omega_{\mu\nu}u^\nu=\frac{nT}{2(e+P)}\nabla_\mu \frac{\mu}{T}+\frac{1}{T}\nabla_\mu T,
\end{align}
where 
 the spin potential is fixed with the thermal vorticity $\omega^{th}_{\mu\nu}$ in accordance with  available conclusions  \cite{Becattini:2012tc,Becattini:2018duy}. On the other hand, a conclusion can be drawn from Eq.(\ref{w0i}) that the components of spin potential parallel with fluid velocity receives the contribution from the combined temperature and chemical potential gradients in equilibrium.
In order to impose the condition of non-negative entropy production, one may cast $\partial_\mu s^\mu$ into a sum of squares like
\begin{align}
\label{square}
ax^2+by^2\geq 0, \quad a,b\geq 0,
\end{align}
which is sufficient for scalar and rank two tensor dissipative processes in the following explicit form
\begin{align}
&
\label{shear}
\pi^{\mu\nu}=2\eta\nabla^{\langle\mu}u^{\nu\rangle}, \\
\label{bulk}
&  \Pi=\zeta\theta,
\\
&
\label{tau}
\tau^{\mu\nu}=2\eta_s \big(\nabla^{[\mu}u^{\nu]}+\Delta^{\mu\rho}\Delta^{\nu\sigma}\omega_{\rho\sigma} \big),\\
\label{tran0}
&\eta\geq 0,\quad\zeta\geq 0,\quad \eta_s\geq 0,
\end{align}
where $\eta$, $\zeta$ and $\eta_s$ represent shear viscosity, bulk viscosity and ``rotational viscosity'' $\eta_s$ \cite{degroot} respectively.
As long as the dissipative quantities take the above form, one can write part of $\partial_\alpha s^\alpha$ as the sum of terms like $ax^2$ or $by^2$. Here $x(y)$ refers to the thermodynamic force, $a(b)$ refers to the correspondent transport coefficient. For example, $ax^2=\zeta\theta^2$ for scalar dissipative process.

However, Eq.(\ref{square}) is a sufficient but not necessary condition if vector dissipative processes are taken into consideration because we can loose the requirement to allow the cross term $xy$ like
\begin{align}
\label{loose}
ax^2+by^2+cxy \geq 0, \quad a,b\geq 0, \quad 4ab\geq c^2.
\end{align}
It can be clearly seen that Eq.(\ref{square}) is a special case of Eq.(\ref{loose}) with $c=0$.
Motivated by this simple illustration,  the sufficient and necessary  conditions of semipositive entropy production is then specifically collected as for vector dissipation quantities:
\begin{eqnarray}
&&
\label{heat}
h^{\mu\prime}=-\kappa\frac{nT}{e+P}\nabla^\mu\frac{\mu}{T}\nn
\\
&&
\quad\quad+\gamma\Big(Du^\mu+\frac{\nabla^\mu T}{T}-2\omega^{\mu\nu}u_\nu\Big),
\\
&&
\label{qu}
q^\mu=\lambda \big(Du^\mu+\frac{\nabla^\mu T}{T}-2\omega^{\mu\nu}u_\nu \big)\nn
\\
&&
\quad\quad+\xi\Big(-\frac{nT}{e+P}\nabla^\mu\frac{\mu}{T}\Big),
\\
&&
\label{tran1}
\kappa\geq 0,\quad\lambda\geq 0, \\
&&
\label{cross}
 \kappa\lambda\geq \frac{1}{4}(\gamma+\xi)^2.
\end{eqnarray}
One can map the linear laws for vectors onto Eq.(\ref{loose}) with
\begin{align}
&x=\frac{n\sqrt{T}}{e+P}\nabla_i\frac{\mu}{T},\\
&y=Du_i+\frac{\nabla_i T}{T}-2\omega_{i\nu}u^\nu,\\
&a=\kappa,\quad b=\lambda,\quad c=\gamma+\xi,
\end{align}
where $i=1,2,3$ denotes spatial indice.
Here  $\kappa$ represent  heat conductivity,  while ``boost heat conductivity'' $\lambda$ \cite{Hattori:2019lfp} together with $\eta_s$ is new in spin hydrodynamics. Moreover,  we propose two new transport coefficients $\gamma$ and $\xi$, which refer to cross effects shown in the vector sector in spin hydrodynamics and  have not appeared in the entropy analysis of related works \cite{Hattori:2019lfp,Fukushima:2020ucl,Hu:2021lnx}. 
An interesting difference from conventional hydrodynamics is that heat conductivity $\kappa$ is bounded from below by $\frac{(\gamma+\xi)^2}{4\lambda}$. This can be regarded as a threshold above which the fluids can sustain heat conduction (the fluctuation with $\kappa$ smaller than $\frac{(\gamma+\xi)^2}{4\lambda}$ is acausal and unstable). 

A cross between rank two tensors, i.e, $\pi^{\mu\nu}$ and $\tau^{\mu\nu}$ is forbidden owing to symmetry.  In essence, the symmetric tensor $\pi^{\mu\nu}$ transforms as a  quintet while  the antisymmetric tensor $\tau^{\mu\nu}$  transforms as a triplet in the Clebsch-Gordan tensor decomposition of SO(3) group $3\otimes 3=3\oplus 5 \oplus 1$, which forbids the cross of $\pi^{\mu\nu}$ and $\tau^{\mu\nu}$.  There is only one representation of SO(3) group for vector  enabling the cross effects  for vector transport. When there is a strong external field, rotational invariance is broken and another cross effects will appear \cite{Cao:2022aku}. There is  still one thing to note. When the fluid on discussion is neutral, then $h^{\mu\prime}\sim O(\partial^2)$,  only one vector dissipative current $q^\mu$ exists and there is no proposed cross effect \cite{Hattori:2019lfp}.

We define $\gamma,\xi$ as the symmetric/antisymmetric cross diffusion coefficients because these cross effects are similar to cross diffusion phenomena widely known in multicharge fluids \cite{Hu:2022vph}.
Noticing that $h^\mu$ appears in the symmetric sector of $T^{\mu\nu}$, then $\gamma$ characterizes the response of $T^{(\mu\nu)}$ to the thermodynamic force belonging to antisymmetric sector $T^{[\mu\nu]}$, vice versa. This can be further interpreted as a reflection of spin-orbit conversion through Eq.(\ref{Scon}).  Therefore the constitutive relations in conventional sector, specifically frame independent current $h^\mu-\frac{n}{e+P}j^\mu$, receive spin corrections, vice versa. In this way, our results show that a chemical-potential gradient can induce an antisymmetric part of the energy-momentum tensor i.e, the four divergence of  the spin tensor, which is closely related to a well known source for polarization named spin Hall effect \cite{Son:2012zy,Chen:2016xtg,Hidaka:2017auj,Yi:2021ryh,Fu:2022myl,Wu:2022mkr}.
Note as an aside,  $\gamma$ should equate with $\xi$ according to Onsager reciprocal relation.


\section{ Non-equilibrium Statistical Operator and Kubo relations} 
 The presence of $\gamma$ and $\xi$  is evident in the language of non-equilibrium statistical operator developed by Zubarev \cite{Zubarev,Hosoya:1983id},
 \begin{eqnarray}
 &\hat{\rho}(t)=Q^{-1}\exp\left[-\int d^3\bx\, \hat{Z}(\bx,t)\right],\\
 &Q=\Tr\exp\left[-\int d^3\bx\, \hat{Z}(\bx,t)\right],
 \end{eqnarray}
 with the operator $\hat{Z}$  defined as
 \begin{align} 
 \label{Z}
 &\hat{Z}(\bx,t)\equiv\epsilon\int_{-\infty}^t dt^\prime e^{\epsilon(t^\prime-t)}
 \Big[\beta^\nu(\bx,t^\prime) \hat{T}_{0\nu}(\bx,t^\prime)
 \nn
 \\
 &-\alpha(\bx,t^\prime)\hat{N}^0(\bx,t^\prime)-\frac{1}{2}\beta(\bx,t^\prime)\omega_{\rho\sigma}(\bx,t^\prime)\hat{S}^{0\rho\sigma}(\bx,t^\prime) \Big],
 \end{align}
 where $\beta$  stands for the
 inverse local  temperature and $\beta^\nu\equiv\beta u^\nu, \alpha\equiv\beta\mu$, and $\epsilon\rightarrow+0$ should be taken after thermodynamic limit.  A generic and natural tensor decomposition reads as:
 \begin{align}
 \label{hatT}
 &
 \hat{T}^{\mu\nu} = \hat{e} u^\mu u^\nu - \hat{p} \Delta^{\mu\nu} + \hat{T}^{(\mu\nu)} +\hat{T}^{[\mu\nu]}
 \, ,
 \\
 \label{hatTs}
 &
 \hat{T}^{(\mu\nu)}= 2\hat{h}^{(\mu}u^{\nu)}+\hat{\pi}^{\mu\nu}+\hat{\Pi}\Delta^{\mu\nu} \, ,
 \\
 \label{hatTa}
 &
 \hat{T}^{[\mu\nu]} = 2\hat{q}^{[\mu}u^{\nu]}+\hat{\tau}^{\mu\nu},
 \, \\
 &\hat{N}^{\mu}=\hat{n}u^\mu+\hat{j}^\mu,\\
 &\hat{S}^{\mu\alpha\beta} =  u^\mu \hat{S}^{\alpha\beta} +\hat{S}^{\mu\alpha\beta} _{(1)},
 \end{align}
 which consistently  matches the form of eqs.(\ref{t}) to (\ref{s}) in hydrodynamic description.
 
 Following the practice detailed in \cite{Hu:2021lnx}, one can work out the linear response of $\hat{T}^{\mu\nu}$ and $\hat{N}^\mu$ to the perturbation. Here we only focus on the vectors $\hat{h}^\prime$ and $\hat{q}$ within them.
Thus the linear thermodynamic current-force relations Eqs.(\ref{heat}) and (\ref{qu}) are reproduced with dissipative quantities replaced by their correspondent operators
and the transport coefficients are  expressed in terms of Kubo correlators:
\begin{align}
\label{kappa} 
\kappa=&\frac{-\beta}{3}\int d^3\bx^{\prime}\int_{-\infty}^0 dt^{\prime}
e^{\epsilon t^{\prime}}\Big( \hat{h}^{\prime\mu}(\bx), \hat{h}_{\mu}^\prime(\bx^{\prime},t^{\prime})\Big),\\
\label{lambda} 
\lambda=&\frac{-\beta}{3}\int d^3\bx^{\prime}\int_{-\infty}^0 dt^{\prime}
e^{\epsilon t^{\prime}}\Big(\hat{q}^\mu(\bx) , \hat{q}_\mu(\bx^{\prime},t^{\prime})\Big),\\
\label{gamma}
\gamma=&\frac{-\beta}{3}\int d^3\bx^{\prime}\int_{-\infty}^0 dt^{\prime}
e^{\epsilon t^{\prime}}\Big(\hat{h}^{\prime\mu}(\bx) , \hat{q}_\mu(\bx^{\prime},t^{\prime})\Big),\\
\label{xi}
\xi=&\frac{-\beta}{3}\int d^3\bx^{\prime}\int_{-\infty}^0 dt^{\prime}
e^{\epsilon t^{\prime}}\Big(\hat{q}^\mu(\bx) , \hat{h}^\prime_\mu(\bx^{\prime},t^{\prime})\Big),
\end{align}
where the Kubo correlator is defined as 
\begin{align}
\label{correlator}
&\left(\hat{X}(\bx,t),\hat{Y}(\bx',t')\right)\equiv\int_0^1 d\tau\Big\langle\hat{X}(\bx,t)\nn\\
&\quad\quad\times\big( e^{-\hat{A}\tau}\hat {Y}(\bx',t')e^{\hat{A}\tau}
-\langle\hat{Y}(\bx',t')\rangle_{l}\big)\Big\rangle_{l}.
\end{align}
 We have taken the ensemble average with the local equilibrium statistical operator
 \begin{align}
 \label{leso} 
 &\hat{\rho}_{l}\equiv Q_{l}^{-1}\exp\Big(-\hat{A}\,\Big),\quad Q_{l}=\Tr\exp\Big(-\hat{A}\,\Big),\\
 &\hat{A}=\int d^3\bx \Big[\beta^\nu(\bx,t)
 \hat{T}_{0\nu}(\bx,t)-\alpha(\bx,t)\hat{N}^0(\bx)
 \nn
 \\
 &\quad
 -\frac{1}{2}\beta(\bx,t)\omega_{\rho\sigma}(\bx,t)\hat{S}^{0\rho\sigma}(\bx,t)\Big].
 \end{align}
 The cross correlations  $(\hat{h}^\prime,\hat{q})$ and $(\hat{q},\hat{h}^\prime)$ appear naturally and no symmetries vanish them.


 Unlike ordinary transport coefficients, e.g, $\kappa$ appearing also in conventional fluids,  new transport coefficients in spin hydrodynamics are not well-defined in strict hydrodynamic limit $\omega\rightarrow 0$. As spin densities are inherently dissipative quantities, they are introduced as dynamic modes in the same fashion as Hydro+ framework \cite{Stephanov:2017ghc}, namely, the frequency scale $\omega\gtrsim \Gamma_s$ ($\Gamma_s$ is the relaxation rate of non-hydrodynamic spin modes). Then the constitutive relations should be replaced by
\begin{equation}
\label{ta1}
T^{[\mu\nu]}=\left\{
	\begin{aligned}
	 & 2q^{[\mu}u^{\nu]}+\tau^{\mu\nu},\quad \Gamma_s\ll \omega\ll \Gamma,\\
	& \,0,\quad\quad\quad \quad\quad\quad\,\omega\ll \Gamma_s.\\
	\end{aligned}
	\right.
\end{equation}
where $\Gamma$ represent the relaxation rate of other non-hydrodynamic modes.
 In strict hydrodynamic limit, spin hydrodynamics reduces to the usual hydrodynamics. In this sense,  these  spin-related transport coefficients should be defined in a constrained limit $\Gamma_s\ll \omega\ll \Gamma$ \cite{Hongo:2021ona},
\begin{align}
&
\label{la}
\lambda=\frac{1}{3}\lim_{\Gamma_s\ll \omega\ll \Gamma}
\frac{\partial}{\partial\omega}\rm{Im} G^R_{\hat{q}\hat{q}}(\b0,\omega),\\
&
\label{ga}
\gamma=\xi=\frac{1}{3}\,\,\lim_{\Gamma_s\ll \omega\ll \Gamma}\,\,
\frac{\partial}{\partial\omega}\rm{Im}G^R_{\hat{h}^\prime\hat{q}}(\b0,\omega),
\end{align}
where $G^R_{\hat{A}\hat{B}}(\omega,\bk)$ is the Fourier transformation of  the retard two point Green function
\begin{align}
 G_{\hat{A}\hat{B}}^R(\bx,t)\equiv-i\theta(t)\left[\hat{A}(\bx,t),\hat{B}({\bf0},0)\right],
\end{align}
and the Onsager relation $\gamma=\xi$ follows directly from
\begin{align}
\label{reversal}
G^R_{\hat{A}\hat{B}}(\bx,t)=G^R_{\hat{B}\hat{A}}(-\bx,t)\eta_A\eta_B,
\end{align}
with $\eta_A$  the parity of $\hat{A}$ under the time reverse ($\hat{h}^\prime$ and $\hat{q}$ have the same parity). 

There are some comments followed in order.
Firstly, analogous to conventional transport coefficients $\eta$, $\zeta$,   $\kappa$, new ones, $\eta_s$ and $\lambda$,  are also defined via self correlation function and thus nonnegative meeting the condition Eqs.(\ref{tran0}) and (\ref{tran1}).
Secondly,  a specific extraction of $\lambda$ from hydrodynamic correlators would lead us to  the same formula as Eq.(\ref{la}). 
Thirdly, the sign of $\gamma$ and $\xi$ is indefinite but they needn't obey nonnegative requirement. 



In the remainder of this section, we  present how to reproduce the entropy analysis by means of the non-equilibrium statistical operator.
Following  \cite{Zubarev}, we define the entropy   as
\begin{align}
S\equiv-\langle\ln\rho_l\rangle_l.
\end{align}
Noticing that $\hat{\rho}$ is a solution  of Liouville's equation when $\epsilon$ is taken to be zero after thermodynamic limit, $S=-\Tr(\hat{\rho}\ln\hat{\rho}))$ is not a qualified 
definition. With the covariant  matching conditions $u_\mu\delta\langle \hat{T}^{\mu\nu}\rangle u_\nu=u_\mu\delta\langle \hat{N}^\mu\rangle=u_\mu\delta\langle \hat{S}^{\mu\rho\sigma}\rangle=0$, the entropy is detailed as  , 
\begin{align}
S=&\ln Q_l+\int d^3\bx \Big[\beta(\bx,t)
\langle\hat{e}(\bx)\rangle-\alpha(\bx,t)\langle\hat{n}(\bx)\rangle
-\frac{1}{2}\beta(\bx,t)\omega_{ij}(\bx,t)\langle\hat{S}^{ij}(\bx)\rangle\Big],
\end{align}
where we have prescribed that the operators in presence are time independent but their expectation values have time dependence following \cite{Zubarev} and we work in the comoving frame $u^\mu=(1,0,0,0)$ for simplicity.

Conveniently,  the change with time of entropy  can be readily expressed as
\begin{align}
\label{ent}
\frac{\partial S}{\partial t}=&\frac{\partial \ln Q_l}{\partial t}+\int d^3\bx \Big[\frac{\partial\beta(\bx,t)}{\partial t}
\langle\hat{e}(\bx)\rangle-\frac{\partial\alpha(\bx,t)}{\partial t}\langle\hat{n}(\bx)\rangle
-\frac{1}{2}\frac{\partial (\beta(\bx,t)\omega_{ij}(\bx,t)\,)}{\partial t}\langle\hat{S}^{ij}(\bx)\rangle\Big]\nn\\
+&\int d^3\bx \Big[\beta(\bx,t)
\frac{\partial \langle\hat{e}(\bx)\rangle}{\partial t}-\alpha(\bx,t)\frac{\partial\langle\hat{n}(\bx)\rangle}{\partial t}
-\frac{1}{2}\beta(\bx,t)\omega_{ij}(\bx,t)\frac{\partial\langle\hat{S}^{ij}(\bx)\rangle}{\partial t}\Big].
\end{align}

One can show that the first line of Eq.(\ref{ent}) vanishes due to nontrivial cancellation.  By utilizing the equations of motions in hydrodynamics and neglecting terms of higher order in gradients,  we obtain
\begin{align}
\label{scon}
\frac{\partial S}{\partial t}
&=-\int d^3\bx \sigma(\bx)-\int d\sigma_i j^i_s(\bx,t)
\end{align}
with the identification of the entropy diffusion flux as the sum of the energy, charge flux and $\langle\hat{q}^i\rangle$
\begin{align}
j_s^i(\bx,t)\equiv\beta(\bx,t)(\langle\hat{h}^i\rangle-\langle\hat{q}^i\rangle)-\alpha(\bx,t)\langle\hat{j}^i\rangle
\end{align}
and $d\sigma^i$ is an element of surface where the fluxes flow in or out.
The entropy production rate   is written as the product of current and force,
\begin{align}
\label{sig}
\sigma\equiv
&\,(\langle\hat{h}^i\rangle-\frac{e+p}{n}\langle\hat{j}^i\rangle\,)\frac{n}{e+p}\nabla_i \alpha+\langle\hat{\Pi}\rangle\theta+\langle\hat{q}^i\rangle(-\beta Du_i+\nabla_i\beta+2\beta\omega_{ij}u^j)\nn\\
&+\langle\hat{\pi}^{ij}\rangle\partial_i u_j+\langle\hat{\tau}^{ij}\rangle(\partial_i u_j+\omega_{ij}),
\end{align}
which is nothing but Eq.(\ref{entropy}) in local rest frame. On the other hand, the spatial integral over Eq.(\ref{sig}) can be cast into a concise self-correlation form as \cite{Zubarev}
\begin{align}
\label{production}
\int d^3\bx \sigma(\bx)=\int dt e^{\epsilon t}\big(C,C(t)\big)\geq 0,
\end{align}
where $C\equiv \sum_a\int d\bx \hat{G}^aF_a$, and $\hat{G}^a$, $F_a$ denote the dissipative quantities in operator form and correspondent thermodynamic force appearing in Eq.(\ref{sig}). They are linearly associated with each other   by the linear laws Eqs.(\ref{shear}) to (\ref{tau}) and Eqs.(\ref{heat}), (\ref{qu}).  Then one can conclude  $\partial_\alpha s^\alpha\ge 0$ has been more rigorously proved by the  method of  statistical operator. As a result, the conditions outlined in Eqs.(\ref{tran0}), (\ref{tran1}) and (\ref{cross}) are automatically satisfied. 


\section{Linear mode analysis}

A  linear mode analysis is given to seek the impacts of cross effects on spin hydrodynamic motion. We choose to perturb the quiescent equilibrium system according to
\begin{align}
\label{perturb}
e (x) &= e_0 + \delta e (x), \quad
P (x) = P_0 + \delta P(x),
\nn
\\
n (x) &= n_0 + \delta n (x), \quad u^i (x) = 0 + \delta u^i (x),
\nn\\
S^{\mu\nu} (x)& = 0 + \delta S^{\mu\nu} (x),\quad\omega^{\mu\nu} (x) = 0 + \delta \omega^{\mu\nu} (x).
\end{align}
in line with the settings of \cite{Hattori:2019lfp}. 
For concreteness,  Landau definition for velocity is taken and $u^{\mu} = (1, \delta u^i) + {\mathcal O}((\delta u)^2)$.  Now $\kappa$ should be interpreted as  the charge diffusion coefficient instead of heat conductivity. 

With the perturbations given in (\ref{perturb}), one can 
linearize the hydrodynamic equations to obtain:
\begin{align}
\label{n0}
0 &=\partial_0\delta n+\bar{n}_0\big(\partial_i\delta\pi^i+ (\lambda_1 -\gamma_1)\partial_i\partial^i \delta e+(\lambda_2 -\gamma_2)\partial_i\partial^i \delta n-(D_b-D_b^\prime)  \partial_i\delta S^{0i} \nn\\
&	-(\xi_1-\kappa_1) \partial_i\partial^i \delta e-(\xi_2-\kappa_2) \partial_i\partial^i \delta n\big)\\
\label{e0}
0 &=
\partial_0\delta e+\partial_i\delta\pi^i+ 2(\lambda_1\partial_i \partial^i \delta e+\lambda_2\partial_i \partial^i \delta n-D_b \partial_i \delta S^{0i} 
\nn\\
&
-\xi_1 \partial_i\partial^i \delta e-\xi_2 \partial_i\partial^i \delta n),
\\
\label{pi}
0 &=
-\partial_0\delta\pi^i-\beta_1\partial_i \delta e-\beta_2\partial_i \delta n+(\eta^\prime+\eta_s^\prime)(\partial_j\partial_j\delta \pi^i-\partial_i\partial_j\delta \pi^j)\nn\\
&+\eta_t\partial_i\partial_j\delta \pi^j+D_s\partial_j\delta S^{ij},
\\
\label{sij}
0 &=
\partial_0 \delta S^{ij}
+ 2 \{ D_s \delta S^{ij}
+ \eta_s' (\partial^i \delta \pi^j - \partial^j \delta \pi^i ) \},
\\
\label{s0i}
0 &=
\partial_0 \delta S^{0i}
- 2(\lambda_1 \partial^i \delta e+\lambda_2 \partial^i \delta n)+2D_b  \delta S^{0i} 	\nn\\
&+2\xi_1 \partial^i \delta e	+2\xi_2 \partial^i \delta n,
\end{align}
where  $\bar{n}_0\equiv\frac{n_0}{e_0+P_0}$ is reduced charge density
and Einstein summation prescription over repeated (spatial) indices is assumed.
In addition,  the following constants are introduced
\begin{align}
&\beta_1
\equiv
\big(\frac{\partial P}{\partial e}\big)_n,\quad \beta_2\equiv\big(\frac{\partial P}{\partial n}\big)_e
,  \quad c_1\equiv\big(\frac{\partial e}{\partial T}\big)_n,\quad c_2\equiv\big(\frac{\partial n}{\partial T}\big)_e,
\nn
\\
&
\chi_s    \equiv \frac{\partial S^{ij}}{   \partial \omega^{ij}  }
, \quad
D_s
\equiv \frac{ 2 \eta_s}{\chi_s},\quad\eta_s' \equiv \frac{\eta_s}{e_0+P_0}
,\nn
\\
&
{\chi_b}   \equiv \frac{\partial S^{i0}}{\partial \omega_{i0}  }, \quad
{D_b}   \equiv \frac{ 2 \lambda}{{\chi_b}},\quad {D_b}^\prime   \equiv \frac{ 2 \gamma}{{\chi_b}}
,
\nn\\
&
\lambda_i\equiv  \frac{\lambda \beta_i }{e_0+P_0}+\frac{\lambda}{c_iT},\quad
\gamma_i\equiv  \frac{\gamma \beta_i }{e_0+P_0}+\frac{\gamma}{c_iT},
\nn\\
&
\xi_i\equiv  \frac{\xi \beta_i }{e_0+P_0}-\frac{\xi}{c_iT},\quad\kappa_i\equiv  \frac{\kappa \beta_i }{e_0+P_0}-\frac{\kappa}{c_iT},
\nn\\
&
\eta_t
\equiv \frac{1}{e_0+P_0} \left( \zeta + \frac{4}{3} \eta \right)
, \quad
\eta^\prime \equiv \frac{\eta}{e_0+P_0}.
\end{align}
One should be cautious that $\chi_b\equiv\frac{\partial S^{i0}}{\partial \omega_{i0}  }=\frac{\beta}{2}\Tr[\hat{\rho}_l(\hat{S}^{i0})^2]\ge 0$ instead of $\chi_b\equiv\frac{\partial S^{i0}}{\partial \omega^{i0}  }$.

For simplicity, we  suppose that  the fluctuations $\delta f=\delta f(t,z)$.  We work in the Fourier space with $\tilde{A}(k)\equiv \int d^4x e^{i(wt-\bm{k}\cdot \bm{x})}A(x)$, and the Fourier transformation of 
linearized hydrodynamic equations eqs.(\ref{n0}) to (\ref{s0i})  will lead us to the  dispersion relations of normal modes in spin hydrodynamics,
\begin{subequations}
	\begin{align}
	\label{ds}
	\omega &= - 2 iD_s, 
	\\
	\label{db}
	\omega &= - 2 i {D_b},
	\\
	\omega   &=
	\begin{cases}
	\displaystyle
	- 2 iD_s - i \eta_s^\prime  k^2 + {\mathcal O} (k^4),
	\\
	- i \eta^\prime  k^2 + O(k^4),
	\label{eta}
	\end{cases}
	\\
	\omega &=
	\begin{cases}
	\displaystyle
	\pm c_s k - i \Gamma_\parallel k^2 + {\mathcal O} (k^3),
	\\
	\displaystyle
	- 2i {D_b} - i D_{b1} k^2 + {\mathcal O} (k^4),
	\\
	\displaystyle
	 -i D k^2 + {\mathcal O} (k^4).	
	\end{cases}
	\label{sound}
	\end{align}
\end{subequations}
where Eqs.(\ref{db}) and (\ref{eta}) are doubly degenerate and 
\begin{align}
\label{coefficient}
c_s^2&\equiv\beta_1+\beta_2\bar{n}_0,\nn\\
\Gamma_\parallel&\equiv \frac{\eta_t}{2}+\frac{\beta_2^2\bar{n}_0^2}{2c_s^2(e_0+P_0)}(\kappa-\frac{\gamma\xi}{\lambda}),\nn\\
D_{b1}&\equiv  \frac{4 \lambda ^2 \chi_{11}+(\gamma-\lambda) ^2 \bar{n}_0^2 \chi_{22}+4 (\gamma\lambda- \lambda ^2 ) \bar{n}_0 \chi_{12}}{\lambda  \left(\chi_{11} \chi_{22}-\chi_{12}^2\right)},\nn\\
D &\equiv \frac{\bar{n}_0^2}{\bar{n}_0^2\chi_{22}-2\bar{n}_0\chi_{12}+\chi_{11}}(\kappa-\frac{\gamma\xi}{\lambda}).
\end{align}

On the other hand, the susceptibility matrix elements are defined as $\chi_{ab}\equiv\frac{\partial\phi_a}{\partial\lambda_b}$ with $\phi=(\delta n,\delta e,\delta \pi^z,\delta S^{0z})$ and $(\delta\mu-\frac{\mu}{T}\delta T,\frac{\delta T}{T},\delta u^z,\delta \omega_{0z})$ \cite{Kovtun:2012rj}. The susceptibility matrix is explicitly written as
\begin{align}
\label{chi}
&\chi_{4\times4}=\nn\\
&\left(
\begin{array}{cccc}
\left(\frac{\partial n}{\partial \mu}\right)_{T}&T \!\left(\frac{\partial n}{\partial T}\right)_{\mu/T} &0&0\\
\left(\frac{\partial e}{\partial \mu}\right)_{T}&T \!\left(\frac{\partial e}{\partial T}\right)_{\mu/T}  & 0 &   0 \\
0&0 & e_0+P_0 & 0 \\
0&0& 0 & \chi_b \\
\end{array}
\right)
\, .
\end{align}

With a straightforward exercise in thermodynamic derivatives,  we have checked that the leading order imaginary parts of these frequencies are negative consistent with theoretical stability requirements stability in the rest frame. However,  the modes may not remain stable in a moving frame, which calls for a detailed calculation. Several comments are followed in order.
 \begin{itemize}
 	\item
 	Besides conventional hydrodynamic modes, namely, two sound modes, one  charge diffusion mode or heat mode, and two shear modes, we find another six non-hydrodynamic gapped modes reflecting non-conserved property of spin densities.  
 	Considering cross effects affect the constitutive relation of the  charge current in Landau frame, the collective motion of spinful fluids should be different from a spinless one. 
    Compared to  conventional hydrodynamics, we have only five hydrodynamic modes in common on which a direct comparison is based. As far as the contributions up to $O(k^2)$ are concerned, only the damping rates of the charge diffusion mode and sound modes receives spin modification of the form $\kappa-\frac{\gamma\xi}{\lambda}$. This is reasonable because  impacts on heat conduction or charge diffusion only have relation with  the sound modes and charge diffusion mode.  The correction originates from the cross diffusion between orbit and spin motion and the cross effects suppress the attenuation of sound modes and charge diffusion mode owing to contrary sign. Note that the suppression factor $\kappa-\frac{\gamma\xi}{\lambda}$ is consistent with Eq.(\ref{cross}) and confirms the conclusion that $\kappa-\frac{\gamma\xi}{\lambda}=0$ is a threshold above which heat mode (or charge diffusion mode) survives.
 	\item
 	Our results can also be readily compared with a closely related work \cite{Hattori:2019lfp} sharing the same settings. Nevertheless, we also take into account the charge diffusion and cross effects and thus find an extra  charge diffusion mode. The leading order damping coefficients of non-hydrodynamic spin modes are all unchanged only $D_{b1}$ receives the correction from cross effects. Viewing that cross effects only source the perturbation of $T^{0i}$ component, the collective motion in $T^{ij}$ sector remain unchanged. When turning off  the charge diffusion and cross effects, our results match the analysis in \cite{Hattori:2019lfp}.

 	\item
 	    In this work, we concentrate our focus on non-conserved spin densities, the motion equations of which are inherently relaxation-type. Consequently, we have not found propagating degrees of freedom in spin sector.  We note that there are associated studies investigating the dispersion relations of hydrodynamic spin modes, where their analysis is carried out for conserved total angular momentum tensor \cite{Hu:2022lpi,Hu:2022mvl}. In that case, 
 	    hydrodynamic spin modes show up and transverse ones propagate with a propagation speed $c_{\text{spin}}$, contrary to what is reported herein. The relation between non-hydrodynamic and hydrodynamic spin modes, and how to map them  deserves further researches.
   \item
   Noticing all non-hydrodynamic spin modes  attenuate according to the damping rate $D_s$ or $D_b$. In the  limit of $k\rightarrow 0$, these gapped modes still decay in contrast to hydrodynamic modes and their presence introduce two new frequency scales $D_s$ and $D_b$. Spin hydrodynamics should be treated as conventional hydrodynamic evolution plus dissipative spin dynamics in a constrained regime $\omega\gtrsim D_s,D_b$.   This is why the constitutive relation Eq.(\ref{ta1}) is supposed to be phenomenologically introduced. After  characteristic time scales $\tau\sim \frac{1}{D_s}$ or $\frac{1}{D_b}$, the system behaviors fall into the control of ordinary spin-averaged hydrodynamics.
 \end{itemize}

As  a supplement,  we stress that the constitutive relations have gone through the test of time reversal  symmetry \cite{Kovtun:2012rj} required by the Onsager relations, manifesting the effectiveness of newly defined  transport coefficients $\eta_s, \lambda,\gamma,\xi$ in spin hydrodynamics.

\section{Summary:}
In this work, we revisit the  first-order spin hydrodynamic theory  from a general entropy analysis and Zubarev's non-equilibrium statistical operator. By carefully rethinking the construction, a new source is found contributing to the constitutive relation of the frame-invariant heat flow or charge diffusion current  $h^\mu-\frac{e+p}{n}j^\mu$ and its correspondence in $T^{[\mu\nu]}$ named $q^\mu$. The new contribution reflects cross effects in spinful fluids originating from the  spin-orbit coupling. Based on the method of non-equilibrium  statistical operator, we show how to identify  these new cross effects and transport coefficients. One interesting finding shows that the transport coefficient $\kappa$ also appearing in conventional fluid dynamics is now bounded from below by nonzero cross transport coefficients.
 A  linear mode analysis  demonstrates that the damping of sound modes and charge diffusion mode (or heat mode) is suppressed by cross effects. These cross effects impact the  motion of $T^{0i}$ and $N^i$, and an immediate extension to the present research is to see that how the analytical solution \cite{Wang:2021wqq,Wang:2021ngp}, or related causality and stability analysis \cite{Daher:2022wzf,Sarwar:2022yzs} will change by rethinking the corrections brought by cross effects. Considering the widespread uses of the  first order hydrodynamic framework, our proposed cross effects should be laid importance on as a nontrivial supplement and reconsidered in some of related studies. 

\section{Acknowledgments:} 
J.H. is grateful to  Xuguang Huang, Shi Pu, Shuzhe Shi, Koichi Hattori and Donglin Wang for helpful discussions. This work was supported by the NSFC Grants No. 11890710, No. 11890712, and No. 12035006.

\begin{appendix}

\end{appendix}

\bibliographystyle{apsrev}
\bibliography{spinmode}

\begin{thebibliography}{53}
\expandafter\ifx\csname natexlab\endcsname\relax\def\natexlab#1{#1}\fi
\expandafter\ifx\csname bibnamefont\endcsname\relax
  \def\bibnamefont#1{#1}\fi
\expandafter\ifx\csname bibfnamefont\endcsname\relax
  \def\bibfnamefont#1{#1}\fi
\expandafter\ifx\csname citenamefont\endcsname\relax
  \def\citenamefont#1{#1}\fi
\expandafter\ifx\csname url\endcsname\relax
  \def\url#1{\texttt{#1}}\fi
\expandafter\ifx\csname urlprefix\endcsname\relax\def\urlprefix{URL }\fi
\providecommand{\bibinfo}[2]{#2}
\providecommand{\eprint}[2][]{\url{#2}}

\bibitem[{\citenamefont{Liang and Wang}(2005)}]{Liang:2004ph}
\bibinfo{author}{\bibfnamefont{Z.-T.} \bibnamefont{Liang}} \bibnamefont{and}
  \bibinfo{author}{\bibfnamefont{X.-N.} \bibnamefont{Wang}},
  \bibinfo{journal}{Phys. Rev. Lett.} \textbf{\bibinfo{volume}{94}},
  \bibinfo{pages}{102301} (\bibinfo{year}{2005}), \bibinfo{note}{[Erratum:
  Phys.Rev.Lett. 96, 039901 (2006)]}, \eprint{nucl-th/0410079}.

\bibitem[{\citenamefont{Voloshin}(2004)}]{Voloshin:2004ha}
\bibinfo{author}{\bibfnamefont{S.~A.} \bibnamefont{Voloshin}}
  (\bibinfo{year}{2004}), \eprint{nucl-th/0410089}.

\bibitem[{\citenamefont{Betz et~al.}(2007)\citenamefont{Betz, Gyulassy, and
  Torrieri}}]{Betz:2007kg}
\bibinfo{author}{\bibfnamefont{B.}~\bibnamefont{Betz}},
  \bibinfo{author}{\bibfnamefont{M.}~\bibnamefont{Gyulassy}}, \bibnamefont{and}
  \bibinfo{author}{\bibfnamefont{G.}~\bibnamefont{Torrieri}},
  \bibinfo{journal}{Phys. Rev. C} \textbf{\bibinfo{volume}{76}},
  \bibinfo{pages}{044901} (\bibinfo{year}{2007}), \eprint{0708.0035}.

\bibitem[{\citenamefont{Becattini et~al.}(2008)\citenamefont{Becattini,
  Piccinini, and Rizzo}}]{Becattini:2007sr}
\bibinfo{author}{\bibfnamefont{F.}~\bibnamefont{Becattini}},
  \bibinfo{author}{\bibfnamefont{F.}~\bibnamefont{Piccinini}},
  \bibnamefont{and} \bibinfo{author}{\bibfnamefont{J.}~\bibnamefont{Rizzo}},
  \bibinfo{journal}{Phys. Rev. C} \textbf{\bibinfo{volume}{77}},
  \bibinfo{pages}{024906} (\bibinfo{year}{2008}), \eprint{0711.1253}.

\bibitem[{\citenamefont{Barnett}(1935)}]{Barnett:1935}
\bibinfo{author}{\bibfnamefont{S.~J.} \bibnamefont{Barnett}},
  \bibinfo{journal}{Rev. Mod. Phys.} \textbf{\bibinfo{volume}{7}},
  \bibinfo{pages}{129} (\bibinfo{year}{1935}).

\bibitem[{\citenamefont{Adamczyk et~al.}(2017)}]{STAR:2017ckg}
\bibinfo{author}{\bibfnamefont{L.}~\bibnamefont{Adamczyk}} \bibnamefont{et~al.}
  (\bibinfo{collaboration}{STAR}), \bibinfo{journal}{Nature}
  \textbf{\bibinfo{volume}{548}}, \bibinfo{pages}{62} (\bibinfo{year}{2017}),
  \eprint{1701.06657}.

\bibitem[{\citenamefont{Alpatov}(2020)}]{Alpatov:2020iev}
\bibinfo{author}{\bibfnamefont{E.}~\bibnamefont{Alpatov}}
  (\bibinfo{collaboration}{), STAR (for the}), \bibinfo{journal}{J. Phys. Conf.
  Ser.} \textbf{\bibinfo{volume}{1690}}, \bibinfo{pages}{012120}
  (\bibinfo{year}{2020}).

\bibitem[{\citenamefont{Adam et~al.}(2019)}]{Adam:2019srw}
\bibinfo{author}{\bibfnamefont{J.}~\bibnamefont{Adam}} \bibnamefont{et~al.}
  (\bibinfo{collaboration}{STAR}), \bibinfo{journal}{Phys. Rev. Lett.}
  \textbf{\bibinfo{volume}{123}}, \bibinfo{pages}{132301}
  (\bibinfo{year}{2019}), \eprint{1905.11917}.

\bibitem[{\citenamefont{Adam et~al.}(2018)}]{Adam:2018ivw}
\bibinfo{author}{\bibfnamefont{J.}~\bibnamefont{Adam}} \bibnamefont{et~al.}
  (\bibinfo{collaboration}{STAR}), \bibinfo{journal}{Phys. Rev. C}
  \textbf{\bibinfo{volume}{98}}, \bibinfo{pages}{014910}
  (\bibinfo{year}{2018}), \eprint{1805.04400}.

\bibitem[{\citenamefont{Hattori et~al.}(2019)\citenamefont{Hattori, Hongo,
  Huang, Matsuo, and Taya}}]{Hattori:2019lfp}
\bibinfo{author}{\bibfnamefont{K.}~\bibnamefont{Hattori}},
  \bibinfo{author}{\bibfnamefont{M.}~\bibnamefont{Hongo}},
  \bibinfo{author}{\bibfnamefont{X.-G.} \bibnamefont{Huang}},
  \bibinfo{author}{\bibfnamefont{M.}~\bibnamefont{Matsuo}}, \bibnamefont{and}
  \bibinfo{author}{\bibfnamefont{H.}~\bibnamefont{Taya}},
  \bibinfo{journal}{Phys. Lett. B} \textbf{\bibinfo{volume}{795}},
  \bibinfo{pages}{100} (\bibinfo{year}{2019}), \eprint{1901.06615}.

\bibitem[{\citenamefont{Fukushima and Pu}(2021)}]{Fukushima:2020ucl}
\bibinfo{author}{\bibfnamefont{K.}~\bibnamefont{Fukushima}} \bibnamefont{and}
  \bibinfo{author}{\bibfnamefont{S.}~\bibnamefont{Pu}}, \bibinfo{journal}{Phys.
  Lett. B} \textbf{\bibinfo{volume}{817}}, \bibinfo{pages}{136346}
  (\bibinfo{year}{2021}), \eprint{2010.01608}.

\bibitem[{\citenamefont{Li et~al.}(2021)\citenamefont{Li, Stephanov, and
  Yee}}]{Li:2020eon}
\bibinfo{author}{\bibfnamefont{S.}~\bibnamefont{Li}},
  \bibinfo{author}{\bibfnamefont{M.~A.} \bibnamefont{Stephanov}},
  \bibnamefont{and} \bibinfo{author}{\bibfnamefont{H.-U.} \bibnamefont{Yee}},
  \bibinfo{journal}{Phys. Rev. Lett.} \textbf{\bibinfo{volume}{127}},
  \bibinfo{pages}{082302} (\bibinfo{year}{2021}), \eprint{2011.12318}.

\bibitem[{\citenamefont{Gallegos et~al.}(2021)\citenamefont{Gallegos, G\"ursoy,
  and Yarom}}]{Gallegos:2021bzp}
\bibinfo{author}{\bibfnamefont{A.~D.} \bibnamefont{Gallegos}},
  \bibinfo{author}{\bibfnamefont{U.}~\bibnamefont{G\"ursoy}}, \bibnamefont{and}
  \bibinfo{author}{\bibfnamefont{A.}~\bibnamefont{Yarom}},
  \bibinfo{journal}{SciPost Phys.} \textbf{\bibinfo{volume}{11}},
  \bibinfo{pages}{041} (\bibinfo{year}{2021}), \eprint{2101.04759}.

\bibitem[{\citenamefont{Florkowski et~al.}(2018)\citenamefont{Florkowski,
  Friman, Jaiswal, and Speranza}}]{Florkowski:2017ruc}
\bibinfo{author}{\bibfnamefont{W.}~\bibnamefont{Florkowski}},
  \bibinfo{author}{\bibfnamefont{B.}~\bibnamefont{Friman}},
  \bibinfo{author}{\bibfnamefont{A.}~\bibnamefont{Jaiswal}}, \bibnamefont{and}
  \bibinfo{author}{\bibfnamefont{E.}~\bibnamefont{Speranza}},
  \bibinfo{journal}{Phys. Rev. C} \textbf{\bibinfo{volume}{97}},
  \bibinfo{pages}{041901} (\bibinfo{year}{2018}), \eprint{1705.00587}.

\bibitem[{\citenamefont{Bhadury et~al.}(2021)\citenamefont{Bhadury, Florkowski,
  Jaiswal, Kumar, and Ryblewski}}]{Bhadury:2020cop}
\bibinfo{author}{\bibfnamefont{S.}~\bibnamefont{Bhadury}},
  \bibinfo{author}{\bibfnamefont{W.}~\bibnamefont{Florkowski}},
  \bibinfo{author}{\bibfnamefont{A.}~\bibnamefont{Jaiswal}},
  \bibinfo{author}{\bibfnamefont{A.}~\bibnamefont{Kumar}}, \bibnamefont{and}
  \bibinfo{author}{\bibfnamefont{R.}~\bibnamefont{Ryblewski}},
  \bibinfo{journal}{Phys. Rev. D} \textbf{\bibinfo{volume}{103}},
  \bibinfo{pages}{014030} (\bibinfo{year}{2021}), \eprint{2008.10976}.

\bibitem[{\citenamefont{Shi et~al.}(2021)\citenamefont{Shi, Gale, and
  Jeon}}]{Shi:2020htn}
\bibinfo{author}{\bibfnamefont{S.}~\bibnamefont{Shi}},
  \bibinfo{author}{\bibfnamefont{C.}~\bibnamefont{Gale}}, \bibnamefont{and}
  \bibinfo{author}{\bibfnamefont{S.}~\bibnamefont{Jeon}},
  \bibinfo{journal}{Phys. Rev. C} \textbf{\bibinfo{volume}{103}},
  \bibinfo{pages}{044906} (\bibinfo{year}{2021}), \eprint{2008.08618}.

\bibitem[{\citenamefont{Peng et~al.}(2021)\citenamefont{Peng, Zhang, Sheng, and
  Wang}}]{Peng:2021ago}
\bibinfo{author}{\bibfnamefont{H.-H.} \bibnamefont{Peng}},
  \bibinfo{author}{\bibfnamefont{J.-J.} \bibnamefont{Zhang}},
  \bibinfo{author}{\bibfnamefont{X.-L.} \bibnamefont{Sheng}}, \bibnamefont{and}
  \bibinfo{author}{\bibfnamefont{Q.}~\bibnamefont{Wang}},
  \bibinfo{journal}{Chin. Phys. Lett.} \textbf{\bibinfo{volume}{38}},
  \bibinfo{pages}{116701} (\bibinfo{year}{2021}), \eprint{2107.00448}.

\bibitem[{\citenamefont{Hu}(2022{\natexlab{a}})}]{Hu:2021pwh}
\bibinfo{author}{\bibfnamefont{J.}~\bibnamefont{Hu}}, \bibinfo{journal}{Phys.
  Rev. D} \textbf{\bibinfo{volume}{105}}, \bibinfo{pages}{076009}
  (\bibinfo{year}{2022}{\natexlab{a}}), \eprint{2111.03571}.

\bibitem[{\citenamefont{Hu}(2022{\natexlab{b}})}]{Hu:2022xjn}
\bibinfo{author}{\bibfnamefont{J.}~\bibnamefont{Hu}}, \bibinfo{journal}{Phys.
  Rev. D} \textbf{\bibinfo{volume}{105}}, \bibinfo{pages}{096021}
  (\bibinfo{year}{2022}{\natexlab{b}}), \eprint{2204.12946}.

\bibitem[{\citenamefont{Weickgenannt et~al.}(2022)\citenamefont{Weickgenannt,
  Wagner, Speranza, and Rischke}}]{Weickgenannt:2022zxs}
\bibinfo{author}{\bibfnamefont{N.}~\bibnamefont{Weickgenannt}},
  \bibinfo{author}{\bibfnamefont{D.}~\bibnamefont{Wagner}},
  \bibinfo{author}{\bibfnamefont{E.}~\bibnamefont{Speranza}}, \bibnamefont{and}
  \bibinfo{author}{\bibfnamefont{D.}~\bibnamefont{Rischke}}
  (\bibinfo{year}{2022}), \eprint{2203.04766}.

\bibitem[{\citenamefont{Hashimoto et~al.}(2015)\citenamefont{Hashimoto, Iizuka,
  and Kimura}}]{Hashimoto:2013bna}
\bibinfo{author}{\bibfnamefont{K.}~\bibnamefont{Hashimoto}},
  \bibinfo{author}{\bibfnamefont{N.}~\bibnamefont{Iizuka}}, \bibnamefont{and}
  \bibinfo{author}{\bibfnamefont{T.}~\bibnamefont{Kimura}},
  \bibinfo{journal}{Phys. Rev. D} \textbf{\bibinfo{volume}{91}},
  \bibinfo{pages}{086003} (\bibinfo{year}{2015}), \eprint{1304.3126}.

\bibitem[{\citenamefont{Garbiso and Kaminski}(2020)}]{Garbiso:2020puw}
\bibinfo{author}{\bibfnamefont{M.}~\bibnamefont{Garbiso}} \bibnamefont{and}
  \bibinfo{author}{\bibfnamefont{M.}~\bibnamefont{Kaminski}},
  \bibinfo{journal}{JHEP} \textbf{\bibinfo{volume}{12}}, \bibinfo{pages}{112}
  (\bibinfo{year}{2020}), \eprint{2007.04345}.

\bibitem[{\citenamefont{Gallegos and G\"ursoy}(2020)}]{Gallegos:2020otk}
\bibinfo{author}{\bibfnamefont{A.~D.} \bibnamefont{Gallegos}} \bibnamefont{and}
  \bibinfo{author}{\bibfnamefont{U.}~\bibnamefont{G\"ursoy}},
  \bibinfo{journal}{JHEP} \textbf{\bibinfo{volume}{11}}, \bibinfo{pages}{151}
  (\bibinfo{year}{2020}), \eprint{2004.05148}.

\bibitem[{\citenamefont{Montenegro et~al.}(2017)\citenamefont{Montenegro,
  Tinti, and Torrieri}}]{Montenegro:2017rbu}
\bibinfo{author}{\bibfnamefont{D.}~\bibnamefont{Montenegro}},
  \bibinfo{author}{\bibfnamefont{L.}~\bibnamefont{Tinti}}, \bibnamefont{and}
  \bibinfo{author}{\bibfnamefont{G.}~\bibnamefont{Torrieri}},
  \bibinfo{journal}{Phys. Rev. D} \textbf{\bibinfo{volume}{96}},
  \bibinfo{pages}{056012} (\bibinfo{year}{2017}), \bibinfo{note}{[Addendum:
  Phys.Rev.D 96, 079901 (2017)]}, \eprint{1701.08263}.

\bibitem[{\citenamefont{Montenegro and Torrieri}(2020)}]{Montenegro:2020paq}
\bibinfo{author}{\bibfnamefont{D.}~\bibnamefont{Montenegro}} \bibnamefont{and}
  \bibinfo{author}{\bibfnamefont{G.}~\bibnamefont{Torrieri}},
  \bibinfo{journal}{Phys. Rev. D} \textbf{\bibinfo{volume}{102}},
  \bibinfo{pages}{036007} (\bibinfo{year}{2020}), \eprint{2004.10195}.

\bibitem[{\citenamefont{Becattini and Tinti}(2010)}]{Becattini:2009wh}
\bibinfo{author}{\bibfnamefont{F.}~\bibnamefont{Becattini}} \bibnamefont{and}
  \bibinfo{author}{\bibfnamefont{L.}~\bibnamefont{Tinti}},
  \bibinfo{journal}{Annals Phys.} \textbf{\bibinfo{volume}{325}},
  \bibinfo{pages}{1566} (\bibinfo{year}{2010}), \eprint{0911.0864}.

\bibitem[{\citenamefont{Becattini and Tinti}(2013)}]{Becattini:2012pp}
\bibinfo{author}{\bibfnamefont{F.}~\bibnamefont{Becattini}} \bibnamefont{and}
  \bibinfo{author}{\bibfnamefont{L.}~\bibnamefont{Tinti}},
  \bibinfo{journal}{Phys. Rev. D} \textbf{\bibinfo{volume}{87}},
  \bibinfo{pages}{025029} (\bibinfo{year}{2013}), \eprint{1209.6212}.

\bibitem[{\citenamefont{Becattini et~al.}(2019)\citenamefont{Becattini,
  Florkowski, and Speranza}}]{Becattini:2018duy}
\bibinfo{author}{\bibfnamefont{F.}~\bibnamefont{Becattini}},
  \bibinfo{author}{\bibfnamefont{W.}~\bibnamefont{Florkowski}},
  \bibnamefont{and} \bibinfo{author}{\bibfnamefont{E.}~\bibnamefont{Speranza}},
  \bibinfo{journal}{Phys. Lett. B} \textbf{\bibinfo{volume}{789}},
  \bibinfo{pages}{419} (\bibinfo{year}{2019}), \eprint{1807.10994}.

\bibitem[{\citenamefont{Hu}(2021)}]{Hu:2021lnx}
\bibinfo{author}{\bibfnamefont{J.}~\bibnamefont{Hu}}, \bibinfo{journal}{Phys.
  Rev. D} \textbf{\bibinfo{volume}{103}}, \bibinfo{pages}{116015}
  (\bibinfo{year}{2021}), \eprint{2101.08440}.

\bibitem[{\citenamefont{Hehl}(1976)}]{Hehl:1976vr}
\bibinfo{author}{\bibfnamefont{F.~W.} \bibnamefont{Hehl}},
  \bibinfo{journal}{Rept. Math. Phys.} \textbf{\bibinfo{volume}{9}},
  \bibinfo{pages}{55} (\bibinfo{year}{1976}).

\bibitem[{\citenamefont{Speranza and Weickgenannt}(2021)}]{Speranza:2020ilk}
\bibinfo{author}{\bibfnamefont{E.}~\bibnamefont{Speranza}} \bibnamefont{and}
  \bibinfo{author}{\bibfnamefont{N.}~\bibnamefont{Weickgenannt}},
  \bibinfo{journal}{Eur. Phys. J. A} \textbf{\bibinfo{volume}{57}},
  \bibinfo{pages}{155} (\bibinfo{year}{2021}), \eprint{2007.00138}.

\bibitem[{\citenamefont{Israel and Stewart}(1979)}]{Israel:1979wp}
\bibinfo{author}{\bibfnamefont{W.}~\bibnamefont{Israel}} \bibnamefont{and}
  \bibinfo{author}{\bibfnamefont{J.}~\bibnamefont{Stewart}},
  \bibinfo{journal}{Annals Phys.} \textbf{\bibinfo{volume}{118}},
  \bibinfo{pages}{341} (\bibinfo{year}{1979}).

\bibitem[{\citenamefont{Becattini}(2012)}]{Becattini:2012tc}
\bibinfo{author}{\bibfnamefont{F.}~\bibnamefont{Becattini}},
  \bibinfo{journal}{Phys. Rev. Lett.} \textbf{\bibinfo{volume}{108}},
  \bibinfo{pages}{244502} (\bibinfo{year}{2012}), \eprint{1201.5278}.

\bibitem[{\citenamefont{de~Groot and Mazur}(2011)}]{degroot}
\bibinfo{author}{\bibfnamefont{S.}~\bibnamefont{de~Groot}} \bibnamefont{and}
  \bibinfo{author}{\bibfnamefont{P.}~\bibnamefont{Mazur}},
  \emph{\bibinfo{title}{Non-Equilibrium Thermodynamics}}
  (\bibinfo{publisher}{Dover Publications, Inc.}, \bibinfo{address}{New York},
  \bibinfo{year}{2011}).

\bibitem[{\citenamefont{Cao et~al.}(2022)\citenamefont{Cao, Hattori, Hongo,
  Huang, and Taya}}]{Cao:2022aku}
\bibinfo{author}{\bibfnamefont{Z.}~\bibnamefont{Cao}},
  \bibinfo{author}{\bibfnamefont{K.}~\bibnamefont{Hattori}},
  \bibinfo{author}{\bibfnamefont{M.}~\bibnamefont{Hongo}},
  \bibinfo{author}{\bibfnamefont{X.-G.} \bibnamefont{Huang}}, \bibnamefont{and}
  \bibinfo{author}{\bibfnamefont{H.}~\bibnamefont{Taya}},
  \bibinfo{journal}{PTEP} \textbf{\bibinfo{volume}{2022}},
  \bibinfo{pages}{071D01} (\bibinfo{year}{2022}), \eprint{2205.08051}.

\bibitem[{\citenamefont{Hu and Shi}(2022)}]{Hu:2022vph}
\bibinfo{author}{\bibfnamefont{J.}~\bibnamefont{Hu}} \bibnamefont{and}
  \bibinfo{author}{\bibfnamefont{S.}~\bibnamefont{Shi}},
  \bibinfo{journal}{Phys. Rev. D} \textbf{\bibinfo{volume}{106}},
  \bibinfo{pages}{014007} (\bibinfo{year}{2022}), \eprint{2204.10100}.

\bibitem[{\citenamefont{Son and Yamamoto}(2013)}]{Son:2012zy}
\bibinfo{author}{\bibfnamefont{D.~T.} \bibnamefont{Son}} \bibnamefont{and}
  \bibinfo{author}{\bibfnamefont{N.}~\bibnamefont{Yamamoto}},
  \bibinfo{journal}{Phys. Rev. D} \textbf{\bibinfo{volume}{87}},
  \bibinfo{pages}{085016} (\bibinfo{year}{2013}), \eprint{1210.8158}.

\bibitem[{\citenamefont{Chen et~al.}(2016)\citenamefont{Chen, Ishii, Pu, and
  Yamamoto}}]{Chen:2016xtg}
\bibinfo{author}{\bibfnamefont{J.-W.} \bibnamefont{Chen}},
  \bibinfo{author}{\bibfnamefont{T.}~\bibnamefont{Ishii}},
  \bibinfo{author}{\bibfnamefont{S.}~\bibnamefont{Pu}}, \bibnamefont{and}
  \bibinfo{author}{\bibfnamefont{N.}~\bibnamefont{Yamamoto}},
  \bibinfo{journal}{Phys. Rev. D} \textbf{\bibinfo{volume}{93}},
  \bibinfo{pages}{125023} (\bibinfo{year}{2016}), \eprint{1603.03620}.

\bibitem[{\citenamefont{Hidaka et~al.}(2018)\citenamefont{Hidaka, Pu, and
  Yang}}]{Hidaka:2017auj}
\bibinfo{author}{\bibfnamefont{Y.}~\bibnamefont{Hidaka}},
  \bibinfo{author}{\bibfnamefont{S.}~\bibnamefont{Pu}}, \bibnamefont{and}
  \bibinfo{author}{\bibfnamefont{D.-L.} \bibnamefont{Yang}},
  \bibinfo{journal}{Phys. Rev. D} \textbf{\bibinfo{volume}{97}},
  \bibinfo{pages}{016004} (\bibinfo{year}{2018}), \eprint{1710.00278}.

\bibitem[{\citenamefont{Yi et~al.}(2021)\citenamefont{Yi, Pu, and
  Yang}}]{Yi:2021ryh}
\bibinfo{author}{\bibfnamefont{C.}~\bibnamefont{Yi}},
  \bibinfo{author}{\bibfnamefont{S.}~\bibnamefont{Pu}}, \bibnamefont{and}
  \bibinfo{author}{\bibfnamefont{D.-L.} \bibnamefont{Yang}},
  \bibinfo{journal}{Phys. Rev. C} \textbf{\bibinfo{volume}{104}},
  \bibinfo{pages}{064901} (\bibinfo{year}{2021}), \eprint{2106.00238}.

\bibitem[{\citenamefont{Fu et~al.}(2022)\citenamefont{Fu, Pang, Song, and
  Yin}}]{Fu:2022myl}
\bibinfo{author}{\bibfnamefont{B.}~\bibnamefont{Fu}},
  \bibinfo{author}{\bibfnamefont{L.}~\bibnamefont{Pang}},
  \bibinfo{author}{\bibfnamefont{H.}~\bibnamefont{Song}}, \bibnamefont{and}
  \bibinfo{author}{\bibfnamefont{Y.}~\bibnamefont{Yin}} (\bibinfo{year}{2022}),
  \eprint{2201.12970}.

\bibitem[{\citenamefont{Wu et~al.}(2022)\citenamefont{Wu, Yi, Qin, and
  Pu}}]{Wu:2022mkr}
\bibinfo{author}{\bibfnamefont{X.-Y.} \bibnamefont{Wu}},
  \bibinfo{author}{\bibfnamefont{C.}~\bibnamefont{Yi}},
  \bibinfo{author}{\bibfnamefont{G.-Y.} \bibnamefont{Qin}}, \bibnamefont{and}
  \bibinfo{author}{\bibfnamefont{S.}~\bibnamefont{Pu}}, \bibinfo{journal}{Phys.
  Rev. C} \textbf{\bibinfo{volume}{105}}, \bibinfo{pages}{064909}
  (\bibinfo{year}{2022}), \eprint{2204.02218}.

\bibitem[{\citenamefont{Zubarev}(1974)}]{Zubarev}
\bibinfo{author}{\bibfnamefont{D.~N.} \bibnamefont{Zubarev}},
  \emph{\bibinfo{title}{Nonequilibrium Statistical Thermodynamics}}
  (\bibinfo{publisher}{Plenum}, \bibinfo{address}{New York},
  \bibinfo{year}{1974}).

\bibitem[{\citenamefont{Hosoya et~al.}(1984)\citenamefont{Hosoya, Sakagami, and
  Takao}}]{Hosoya:1983id}
\bibinfo{author}{\bibfnamefont{A.}~\bibnamefont{Hosoya}},
  \bibinfo{author}{\bibfnamefont{M.-a.} \bibnamefont{Sakagami}},
  \bibnamefont{and} \bibinfo{author}{\bibfnamefont{M.}~\bibnamefont{Takao}},
  \bibinfo{journal}{Annals Phys.} \textbf{\bibinfo{volume}{154}},
  \bibinfo{pages}{229} (\bibinfo{year}{1984}).

\bibitem[{\citenamefont{Stephanov and Yin}(2018)}]{Stephanov:2017ghc}
\bibinfo{author}{\bibfnamefont{M.}~\bibnamefont{Stephanov}} \bibnamefont{and}
  \bibinfo{author}{\bibfnamefont{Y.}~\bibnamefont{Yin}},
  \bibinfo{journal}{Phys. Rev. D} \textbf{\bibinfo{volume}{98}},
  \bibinfo{pages}{036006} (\bibinfo{year}{2018}), \eprint{1712.10305}.

\bibitem[{\citenamefont{Hongo et~al.}(2021)\citenamefont{Hongo, Huang,
  Kaminski, Stephanov, and Yee}}]{Hongo:2021ona}
\bibinfo{author}{\bibfnamefont{M.}~\bibnamefont{Hongo}},
  \bibinfo{author}{\bibfnamefont{X.-G.} \bibnamefont{Huang}},
  \bibinfo{author}{\bibfnamefont{M.}~\bibnamefont{Kaminski}},
  \bibinfo{author}{\bibfnamefont{M.}~\bibnamefont{Stephanov}},
  \bibnamefont{and} \bibinfo{author}{\bibfnamefont{H.-U.} \bibnamefont{Yee}},
  \bibinfo{journal}{JHEP} \textbf{\bibinfo{volume}{11}}, \bibinfo{pages}{150}
  (\bibinfo{year}{2021}), \eprint{2107.14231}.

\bibitem[{\citenamefont{Kovtun}(2012)}]{Kovtun:2012rj}
\bibinfo{author}{\bibfnamefont{P.}~\bibnamefont{Kovtun}}, \bibinfo{journal}{J.
  Phys. A} \textbf{\bibinfo{volume}{45}}, \bibinfo{pages}{473001}
  (\bibinfo{year}{2012}), \eprint{1205.5040}.

\bibitem[{\citenamefont{Hu}(2022{\natexlab{c}})}]{Hu:2022lpi}
\bibinfo{author}{\bibfnamefont{J.}~\bibnamefont{Hu}}, \bibinfo{journal}{Phys.
  Rev. D} \textbf{\bibinfo{volume}{106}}, \bibinfo{pages}{036004}
  (\bibinfo{year}{2022}{\natexlab{c}}), \eprint{2202.07373}.

\bibitem[{\citenamefont{Hu and Xu}(2022)}]{Hu:2022mvl}
\bibinfo{author}{\bibfnamefont{J.}~\bibnamefont{Hu}} \bibnamefont{and}
  \bibinfo{author}{\bibfnamefont{Z.}~\bibnamefont{Xu}} (\bibinfo{year}{2022}),
  \eprint{2205.15755}.

\bibitem[{\citenamefont{Wang et~al.}(2022)\citenamefont{Wang, Xie, Fang, and
  Pu}}]{Wang:2021wqq}
\bibinfo{author}{\bibfnamefont{D.-L.} \bibnamefont{Wang}},
  \bibinfo{author}{\bibfnamefont{X.-Q.} \bibnamefont{Xie}},
  \bibinfo{author}{\bibfnamefont{S.}~\bibnamefont{Fang}}, \bibnamefont{and}
  \bibinfo{author}{\bibfnamefont{S.}~\bibnamefont{Pu}}, \bibinfo{journal}{Phys.
  Rev. D} \textbf{\bibinfo{volume}{105}}, \bibinfo{pages}{114050}
  (\bibinfo{year}{2022}), \eprint{2112.15535}.

\bibitem[{\citenamefont{Wang et~al.}(2021)\citenamefont{Wang, Fang, and
  Pu}}]{Wang:2021ngp}
\bibinfo{author}{\bibfnamefont{D.-L.} \bibnamefont{Wang}},
  \bibinfo{author}{\bibfnamefont{S.}~\bibnamefont{Fang}}, \bibnamefont{and}
  \bibinfo{author}{\bibfnamefont{S.}~\bibnamefont{Pu}}, \bibinfo{journal}{Phys.
  Rev. D} \textbf{\bibinfo{volume}{104}}, \bibinfo{pages}{114043}
  (\bibinfo{year}{2021}), \eprint{2107.11726}.

\bibitem[{\citenamefont{Daher et~al.}(2022)\citenamefont{Daher, Das, and
  Ryblewski}}]{Daher:2022wzf}
\bibinfo{author}{\bibfnamefont{A.}~\bibnamefont{Daher}},
  \bibinfo{author}{\bibfnamefont{A.}~\bibnamefont{Das}}, \bibnamefont{and}
  \bibinfo{author}{\bibfnamefont{R.}~\bibnamefont{Ryblewski}}
  (\bibinfo{year}{2022}), \eprint{2209.10460}.

\bibitem[{\citenamefont{Sarwar et~al.}(2022)\citenamefont{Sarwar, Hasanujjaman,
  Bhatt, Mishra, and Alam}}]{Sarwar:2022yzs}
\bibinfo{author}{\bibfnamefont{G.}~\bibnamefont{Sarwar}},
  \bibinfo{author}{\bibfnamefont{M.}~\bibnamefont{Hasanujjaman}},
  \bibinfo{author}{\bibfnamefont{J.~R.} \bibnamefont{Bhatt}},
  \bibinfo{author}{\bibfnamefont{H.}~\bibnamefont{Mishra}}, \bibnamefont{and}
  \bibinfo{author}{\bibfnamefont{J.-e.} \bibnamefont{Alam}}
  (\bibinfo{year}{2022}), \eprint{2209.08652}.

\end{thebibliography}

\end{document}